\documentclass[12pt]{article} 
\usepackage{graphicx}
\usepackage{epsfig}
\usepackage{amsbsy}

\begin{document}
\baselineskip 10mm

\centerline{\large \bf Numerical simulation of the thermal fragmentation process in fullerene C$_{60}$}

\vskip 6mm

\centerline{L. A. Openov$^{*}$ and A. I. Podlivaev}

\vskip 4mm

\centerline{\it Moscow Engineering Physics Institute (State
University), 115409 Moscow, Russia}

\vskip 2mm

$^{*}$ E-mail: LAOpenov@mephi.ru

\vskip 8mm

\centerline{\bf ABSTRACT}

The processes of defect formation and annealing in fullerene
C$_{60}$ at $T$ = (4000 - 6000) K are studied by the molecular
dynamics technique with a tight-binding potential. The cluster
lifetime until fragmentation due to the loss of a C$_2$ dimer has
been calculated as a function of temperature. The activation
energy and the frequency factor in the Arrhenius equation for the
fragmentation rate have been found to be $E_a=(9.2\pm 0.4)$ eV and
$A=(8\pm 1)\cdot 10^{19}$ s$^{-1}$. It is shown that fragmentation
can occur after the C$_{60}$ cluster loses its spherical shape.
This fact must be taken into account in theoretical calculations
of $E_a$.

\newpage

In spite of intensive experimental and theoretical studies of
fullerene C$_{60}$ [1], there is still controversy concerning some
of its physical and chemical characteristics. In particular, a
significant discrepancy between theory and experiment in the
determination of the energy $D$ required for the loss of a C$_2$
dimer from fullerene C$_{60}$ was repeatedly noted until the late
1990s. Calculations performed using the most accurate {\it ab
initio} methods gave the value $D$ = (11 - 12) eV [2] (see also
one of the most recent theoretical works [3]), which exceeded the
experimental value of $D$ = (5 - 8) eV by a factor of 1.5 - 2; see
references in [4]. Note, however, that the fragmentation
(dissociation) energy was calculated theoretically by the equation
\begin{equation}
D=E(\textnormal{C}_{58})+E(\textnormal{C}_{2})-E(\textnormal{C}_{60}),
\label{D}
\end{equation}
where $E(C_n)$ is the energy of the corresponding clusters, while $D$ in the analysis of experimental data was usually
assumed equal to the activation energy $E_a$ [4] entering into the Arrhenius equation for the fragmentation rate
\begin{equation}
k(T)=A\cdot\textnormal{exp}(-E_{a}/k_{B}T),
\label{k}
\end{equation}
where $k_{B}$ is the Boltzmann constant, $T$ is the temperature,
and $A$ is the frequency factor in s$^{-1}$. It should be noted
here that the fragmentation energy $D$ and the activation energy
$E_a$, generally speaking, have a different physical meaning;
therefore, their comparison is not quite straightforward.

When $E_a$ was determined by Eq. (2), it was commonly assumed that
$A \sim 10^{15}$ s$^{-1}$ as in the majority of mid-size clusters.
More recently, it has been shown in a series of publications [4 -
7] that the value of $A$ for fullerene C$_{60}$ can be much higher
than had been considered previously, $A \sim (10^{19} - 10^{20})$
s$^{-1}$. Assuming that $A = 5\cdot 10^{19}$ s$^{-1}$, the authors
of [7] reanalyzed the results of a great number of experiments and
came to the conclusion that $E_a = (10.5\pm 0.2)$ eV, where the
arithmetic mean and the standard error were calculated with
"improved" experimental values. This new "corrected" value of
$E_a$ is in much better agreement with both the theoretical value
of $D$ [2, 3] and the more recent experimental data [8 - 13].
However, when revising the experiments carried out in the past
years, the authors of [7], for various reasons, excluded from
their statistical treatment a number of publications  in which the
activation energy $E_a$ proved to be substantially lower than 10
eV even after its recalculation with a different frequency factor.
Moreover, the choice of the value $A = 5\cdot 10^{19}$ s$^{-1}$ in
[7] (or $A = 2\cdot 10^{19}$ s$^{-1}$ and even $A = 3.4\cdot
10^{21}$ s$^{-1}$ in the subsequent publications [9, 10, 12])
seems to be somewhat artificial. This leaves a certain feeling of
disappointment and looks like an attempt to fit the experiment to
theory.

It is also appropriate to note the following circumstance. In the
calculation of the fragmentation energy $D$ of fullerene C$_{60}$
by Eq. (1), it was suggested that both the initial cluster and its
dissociation product (C$_{58}$ cluster) are in their optimal (most
energy-favorable) configurations with the lowest possible energy
$E(C_n$) at a given number of carbon atoms in the cluster ($n$ =
60 or 58). This corresponds to so-called "adiabatic dissociation".
At the same time, under real conditions, the C$_2$ dimer can be
detached from a C$_{60}$ cluster which is (at least) strongly
distorted due to an external action and/or thermal vibrations
rather than from the perfect fullerene. The structure of the
C$_{58}$ cluster remaining after the evaporation of the C$_2$
dimer can also differ from the most stable one (standard
mass-spectrometry methods can determine the number of atoms in a
cluster but not its shape).

From all the above, it follows that a real-time numerical simulation of fullerene C$_{60}$ fragmentation is
of great interest. It allows the temperature dependence of the cluster lifetime
\begin{equation}
\tau(T)=\frac{1}{k(T)}
\label{tau}
\end{equation}
to be determined directly and the values of $A$ and $E_a$ to be
found from Eq. (2) without any additional assumptions. The results
of simulations of fullerene C$_{60}$ dynamics at high temperature
have been presented previously in a number of works. In this case,
however, the main attention has been paid to studying "phase
transformations" upon heating the cluster [14 - 16] or the
mechanism of C$_2$ dimer loss from the cluster [17 - 19]. In the
present work, all the steps of fullerene C$_{60}$ fragmentation
are studied, starting with the formation of defect isomers and
ending with the loss of a C$_2$ dimer. We used tight-binding
molecular dynamics [20], in which the many-body tight-binding
potential allowed the electronic subsystem contribution to the
total energy to be determined more accurately than in the case of
simple classical potentials. This potential did not require such a
large expenditure of computer resources as {\it ab initio}
approaches, thus making it possible to follow the evolution of
fullerene C$_{60}$ during $t \approx 20$ ns (while usually $t <.1$
ns in {\it ab initio} calculations). At the same time, for bond
lengths, binding energies, the energy gap between the highest
occupied and lowest unoccupied molecular orbitals (HOMO-LUMO gap),
potential barrier heights in the Stone-Wales transformation [21],
and other characteristics of fullerene C$_{60}$, the tight-binding
potential gives values that are in good agreement with
experimental data and/or {\it ab initio} calculations (for more
details, see [19]). Using Eq. (1) and this potential, we
calculated the most energy-favorable configurations of C$_{58}$
and C$_{60}$ clusters and obtained $D$ = 11.04 eV, which is also
in agreement with theoretical data available in the literature [2,
3].

The numerical simulation of fullerene C$_{60}$ dynamics was
performed at a fixed total energy of the cluster [22]. This
approach to the problem corresponds to a real experimental
situation in the fragmentation of fullerenes when clusters excited
by a laser pulse do not collide with each other during the time
$t$ \textless{} 1 $\mu$s required for their fragmentation; that
is, they are not in thermal equilibrium with the environment (see
[17]). In this case, the cluster temperature $T$ is a measure of
the relative motion of its constituent atoms, that is, a measure
of the excitation energy after irradiation with a laser. It was
determined by the equation $\frac{1}{2}k_{B}T(3n-6)=\langle
E_{\textnormal{kin}}\rangle$, where $n=60$, and $\langle
E_{\textnormal{kin}}\rangle$ is the kinetic energy averaged over
10$^3$ - 10$^4$ molecular-dynamics steps (the time of one step is
$t_0$ = 0.272 fs). At the initial time, random velocities and
displacements were imparted to all the atoms so that the total
momentum and angular momentum were equal to zero. The forces
acting on the atoms were calculated by the Hellmann-Feynman
theorem using the tight-binding Hamiltonian. After that, the
classical Newtonian equations of motion were solved numerically.

Since the total energy, that is, the sum of the kinetic and
potential ($E_{pot}$) energies, remains constant in the process of
cluster evolution, its temperature $T$ (as a measure of the
kinetic energy) decreases when the cluster transforms to a
metastable state with a higher value of $E_{pot}$ [a lower binding
energy $E_b = 60E_{pot}(1) - E_{pot}(60)$] and increases in the
opposite case. The dependence of $T$ on time $t$ is presented in
Fig. 1a at the initial temperature $T_{ini} = (4250 \pm 15)$ K.
The dependence $T(t)$ exhibits clearly defined steps. They are
associated with the formation and annealing of defect isomers
obtained from fullerene C$_{60}$ and from each other as a result
of the so-called Stone-Wales transformation [23], which represents
a transposition of two C-C bonds. While carbon atoms in initial
fullerene C$_{60}$ are located at the vertices of twenty hexagons
and twelve pentagons isolated from each other, the first
Stone-Wales transformation at $t \approx 2.2$ ns gives an isomer
containing two pairs of pentagons with common sides. Its potential
energy is higher than that of fullerene C$_{60}$ by $\Delta
E_{pot} = 1.42$ eV [19]. As a consequence, the cluster temperature
drops stepwise by $\Delta T \approx 100$ K. If each pair of
neighboring pentagons is considered as one "5/5 defect," then this
isomer contains $N_{5/5}$ = 2 such defects (for other defect
isomers, the value of $E_{pot}$ is higher; no isomer with $N_{5/5}
= 1$ exists for topological reasons).

At $t \approx 2.7$ ns, the temperature drops again by $\Delta T
\approx 100$ K. An analysis of the structure of the cluster showed
that an isomer with $N_{5/5} = 4$ forms in this case. Then (at
$t\approx 3.1$ ns), one defect and, almost immediately, another
one are annealed. As a result, an isomer with $N_{5/5} = 2$ forms
again. After annealing of all defects, the structure of the
initial fullerene C$_{60}$ is restored at $t \approx 4.4$ ns. This
structure is retained up to $t \approx 7.6$ ns, after which a
sequence of transitions starts, leading eventually to the loss of
the C$_2$ dimer at $t \approx 9.5$ ns; see Fig. 2a. Note that a
nonclassical fullerene forms at $t \approx 7.9$ ns that contains
one heptagon along with penta- and hexagons, and tetragons appear
subsequently. The C$_{58}$ cluster that forms after fragmentation
also represents a metastable isomer of fullerene C$_{58}$. It
contains $N_{5/5} = 6$ pairs of neighboring pentagons, while
$N_{5/5} = 3$ in fullerene C$_{58}$ with the lowest possible
energy.

Though the C$_2$ dimer in the above example of fullerene C$_{60}$
fragmentation is detached from a strongly defect C$_{60}$ cluster,
this cluster still has a closed spherical "shell;" see Fig. 2a.
The C$_{58}$ cluster forming as a result of fragmentation also has
the same shape. However, our analysis showed that this is not
necessarily the case. We will give an alternative example. The
dependence $T(t)$ at $T_{ini} = (4335 \pm 10)$ K is presented in
Fig. 1b. The narrow deep minimum in $T(t)$ at $t \approx 1.6$ ns
is associated with the formation and rapid annealing (within
$\Delta t \approx 0.1$ ns) of an isomer containing a dodecagon
along with a heptagon. Actually, the cluster in this case "opens"
for a short time, and a large "window" appears on its surface. The
same occurs at $t \approx 3.6$ and at $t \approx 4.8$ ns. The next
time, such a window forms at $t \approx 5.2$ ns; however, now it
adjoins two heptagons rather than one. In addition, the cluster
also has two pairs of neighboring pentagons. The closed surface is
not restored up to the loss of the C$_2$ dimer from the cluster.
The C$_{58}$ cluster remaining after fragmentation is also
"opened," see Fig. 2b.

Since the fragmentation of fullerene C$_{60}$ is of stochastic
character, it is reasonable to consider the mean lifetime $\tau$
at a certain temperature rather than the "fragmentation
temperature" [24]. In order to collect statistical data sufficient
for determining the temperature dependence of $\tau$, we performed
calculations at several tens of values of $T_{ini} = 4000 - 6000$
K corresponding to various sets of initial velocities and
displacements. We found correlation between $T_{ini}$ and the
character of the C$_{60}$ cluster distortion immediately before
fragmentation. The cluster before fragmentation is more frequently
opened than closed at $T_{ini} > 5000$ K and the other way around
at $T_{ini} < 5000$ K. Although the fragmentation proceeded by
means of the loss of a C$_2$ dimer in the majority of cases, the
evaporation of either one carbon atom, or a C$_3$ trimer, or a
C$_4$ tetramer occurred several times, as was also observed
experimentally [25]. With the aim of comparing the results of
simulations with experiments on the loss of a C$_2$ dimer, we
retained for analysis only variants in which fragmentation
proceeded through the channel
$\textnormal{C}_{60}\rightarrow\textnormal{C}_{58}+\textnormal{C}_{2}$.
In order to take into account the effects due to the small cluster
size and the absence of heat exchange with the environment, we
used the following equation for the fragmentation rate instead of
Eq. (2):
\begin{equation}
k(T)=A\cdot\textnormal{exp}\left[-\frac{E_{a}}{k_{B}T(1-\frac{E_{a}}{2CT})}\right].
\label{k2}
\end{equation}
This equation is derived from the finite-heat-bath theory in the
first-order approximation with respect to the small parameter
$E_a/2CT$ [26, 27]. Here, $C$ is the microcanonical heat capacity
of the cluster. We assumed that $C = (3n - 6)k_B$, where $n = 60$.
Since $E_a/2CT \approx 0.07$ at $T \approx 5000$ K, the possible
difference between $C$ and $(3n - 6)k_B$ does not lead to any
significant error in the determination of $E_a$ from Eq. (4).

The calculated dependence of the logarithm of the fullerene
lifetime $\tau = 1/k$ on is presented in Fig. 3, where $T_{ini}^*
= T_{ini}(1 - E_a/2CT_{ini})$, see Eq. 4. In a rather wide range
of $T_{ini} = (4000 - 6000)$ K and, correspondingly, $\tau = 1$ ps
- 20 ns, the results of numerical simulation are described well by
Eq. (4) with $A$ = const($T$). The statistical scatter of the data
prevents the temperature dependence of $A$ from being reliably
determined. However, it may be argued, in any case, that this
dependence if sufficiently weak in this range of $T_{ini}$, since
the dependence of $ln(\tau)$ on $(T_{ini}^*)^{-1}$ is approximated
well by a straight line, see Fig. 3. The activation energy $E_a$
and the frequency factor $A$ can be determined by the slope of
this line and the point of its intersection with the ordinate
axis. For the mean values and standard errors of these quantities,
we obtained $E_a = (9.2 \pm 0.4)$ eV and $A = (8 \pm 1)\cdot
10^{19}$ s$^{-1}$ (since $E_a$ enters both into the numerator of
the exponent in Eq. (4) and into the "renormalized" initial
temperature, it was determined self-consistently by successive
iterations).

Previously, ideas of the loss of a C$_2$ dimer from a strongly
defect C$_{60}$ cluster were discussed as a hypothesis for the
explanation of the discrepancy between the values of $D$ and $E_a$
[25]. However, these considerations were not confirmed by the
corresponding calculations (though evidence for the two-step
character of fullerene C$_{60}$ fragmentation was obtained in [28]
using the semiempirical Tersoff potential). Note that the value of
$E_a$ found in our work is somewhat lower than the averaged
experimental value $E_a = (10.5 \pm 0.2)$ eV [7]. However, if the
experiments giving $E_a < 10$ eV were not rejected, as was done by
the authors of [7], the "experimental mean" value $E_a = (8 - 9)$
eV would almost coincide with our result. We emphasize the fact
that the value of the frequency factor $A$ found in our work by
direct calculations without any assumptions is in reasonable
agreement with the values $A = (2 - 5)\cdot 10^{19}$ s$^{-1}$,
which have been used in recent years by the majority of authors in
the analysis of experimental data [7, 9, 11-13], and does not
strongly differ from the value $A \sim 10^{21}$ s$^{-1}$ reported
in the most recent publication in this field [29].

Note in conclusion that it is interesting to calculate the
lifetime of fullerene C$_{60}$ at lower temperatures $T = (1000 -
4000)$ K and possibly to determine the temperature dependence of
the frequency factor. With molecular dynamics, this task, however,
is beyond the possibilities of the present-day computational
technique. For its solution, fundamentally different approaches
are required based, for example, on Monte Carlo algorithms.

\newpage
\centerline{Figure captions}

\vskip 5mm Fig. 1. Time dependence of the fullerene C$_{60}$
temperature obtained by numerical molecular dynamics simulations.
The total energy of the cluster is constant, and the time of one
step is $t_0$ = 0.272 fs. The values of $T$ were calculated by
averaging the kinetic energy over 10$^4$ steps. The initial
temperature $T_{ini}$ = (a) $(4250 \pm 15)$ K and (b) $(4335 \pm
10)$ K. The sequence of the formation and annealing of defect
C$_{60}$ isomers leading to the loss of a C$_2$ dimer is as
follows: $\textnormal{C}_{60}\rightarrow$
$\textnormal{C}_{60}(N_{5/5}=2)\rightarrow$
$\textnormal{C}_{60}(N_{5/5}=4)\rightarrow$
$\textnormal{C}_{60}(N_{5/5}=3)\rightarrow$
$\textnormal{C}_{60}(N_{5/5}=4)\rightarrow$
$\textnormal{C}_{60}(N_{5/5}=2)\rightarrow$
$\textnormal{C}_{60}\rightarrow$
$\textnormal{C}_{60}(N_{5/5}=2)\rightarrow$
$\textnormal{C}_{60}(N_{5/5}=4,N_{7}=1)\rightarrow$
$\textnormal{C}_{60}(N_{5/5}=6,N_{7}=1)\rightarrow$
$\textnormal{C}_{60}(N_{5/5}=7,N_{7}=1)\rightarrow$
$\textnormal{C}_{60}(N_{5/5}=9,N_{7}=2)\rightarrow$
$\textnormal{C}_{60}(N_{5/5}=4,N_{7}=1,N_{4}=1)\rightarrow$
$\textnormal{C}_{60}(N_{5/5}=6)\rightarrow$
$\textnormal{C}_{60}(N_{5/5}=4,N_{7}=1,N_{4}=1)\rightarrow$
$\textnormal{C}_{58}+\textnormal{C}_{2}$ for (a) and
$\textnormal{C}_{60}\rightarrow$
$\textnormal{C}_{60}(N_{5/5}=2)\rightarrow$
$\textnormal{C}_{60}\rightarrow$
$\textnormal{C}_{60}(N_{7}=1,N_{12}=1)\rightarrow$
$\textnormal{C}_{60}\rightarrow$
$\textnormal{C}_{60}(N_{7}=1,N_{12}=1)\rightarrow$
$\textnormal{C}_{60}\rightarrow$
$\textnormal{C}_{60}(N_{5/5}=2)\rightarrow$
$\textnormal{C}_{60}(N_{5/5}=4,N_{7}=1)\rightarrow$
$\textnormal{C}_{60}\rightarrow$
$\textnormal{C}_{60}(N_{5/5}=4,N_{7}=1)\rightarrow$
$\textnormal{C}_{60}(N_{5/5}=2)\rightarrow$
$\textnormal{C}_{60}(N_{5/5}=5,N_{7}=1)\rightarrow$
$\textnormal{C}_{60}(N_{5/5}=4,N_{7}=2,N_{12}=1)\rightarrow$
$\textnormal{C}_{58}+\textnormal{C}_{2}$ for (b), where $N_{5/5}$,
$N_4$, $N_7$, and $N_{12}$ are the numbers of pairs of pentagons
with common sides, tetragons, heptagons, and dodecagons,
respectively; see text.

Fig. 2. Snapshots of a C$_{60}$ cluster before and immediately
after fragmentation. The time interval between two snapshots
is $\Delta t\approx 2$ ps. For clarity, the distant atoms are not
shown. The atoms removed upon fragmentation are shown in
black. The temperature prehistory of fragmentation processes (a)
and (b) are given in Figs. 1a and 1b, respectively.

Fig. 3. Logarithm of the lifetime $\tau$ (in seconds) of fullerene
C$_{60}$ before $\textnormal{C}_{60}\rightarrow\textnormal{C}_{58}+\textnormal{C}_{2}$
fragmentation vs. the reciprocal initial temperature (kelvins) including the finite-heat-bath
correction, see text. The circles are the results of calculations and the solid line is a linear least-squares approximation.

\newpage
\includegraphics[width=9cm,height=19cm]{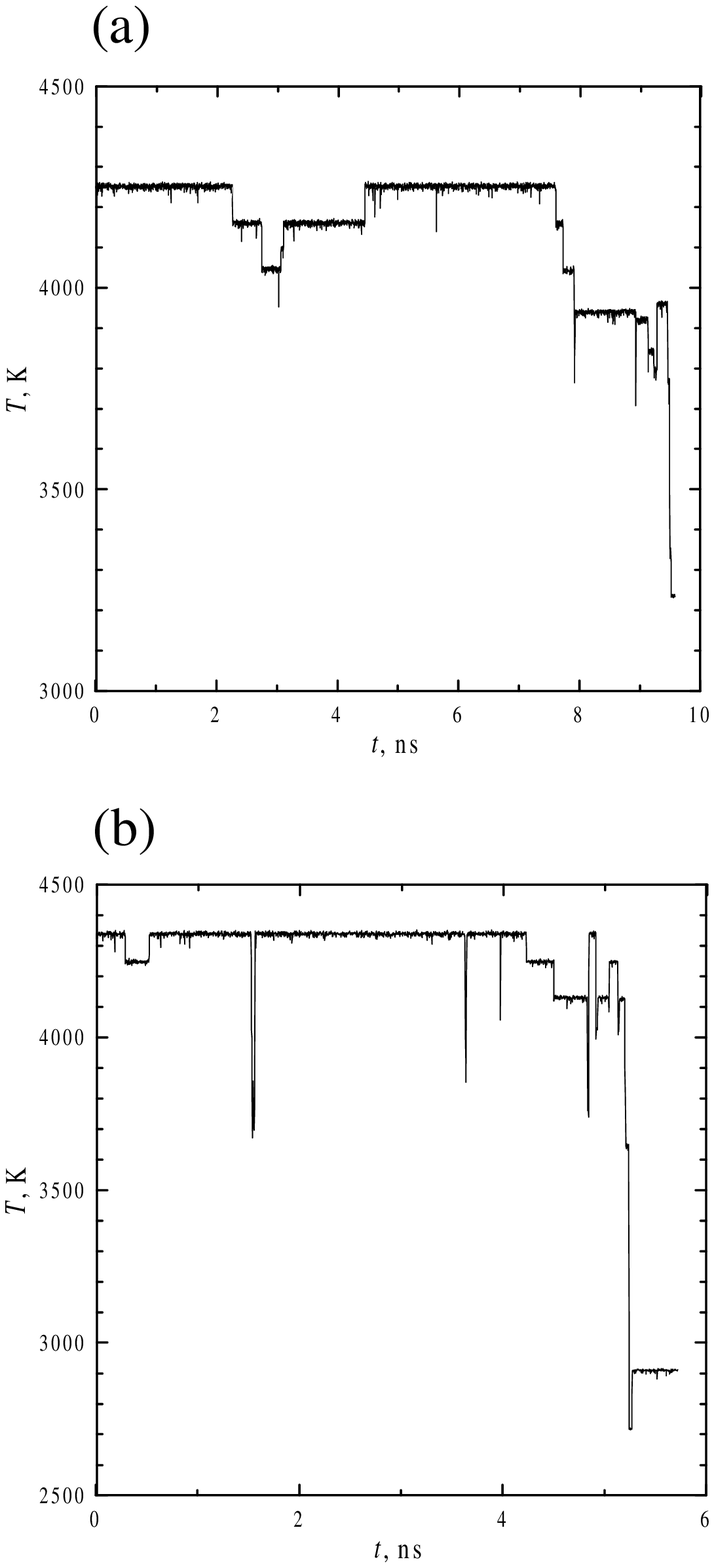}
\vskip 20mm \centerline{Figure 1}

\newpage
\includegraphics[width=15cm,height=\hsize]{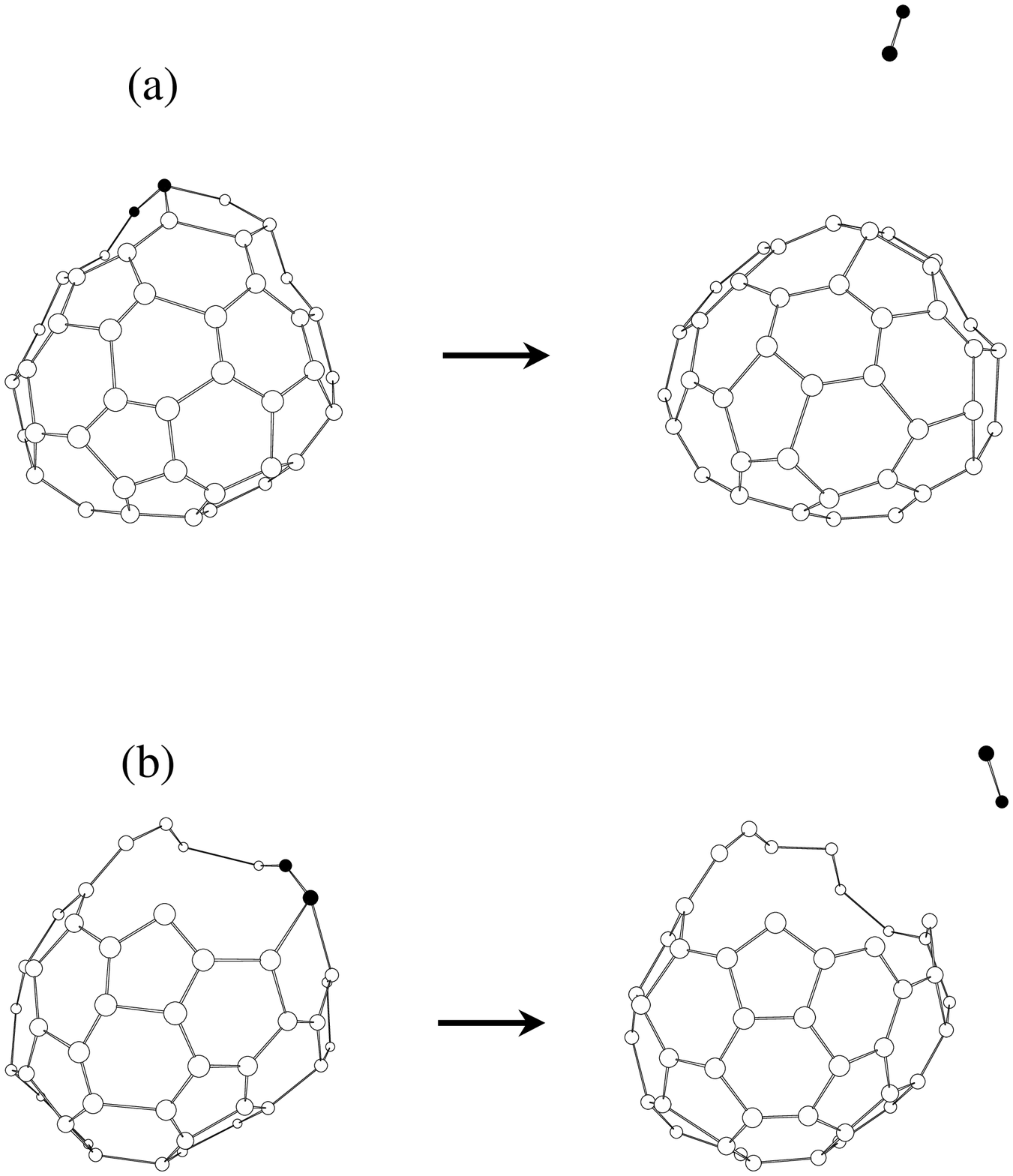}
\vskip 20mm
\centerline{Figure 2}

\newpage
\includegraphics[width=15cm,height=14cm]{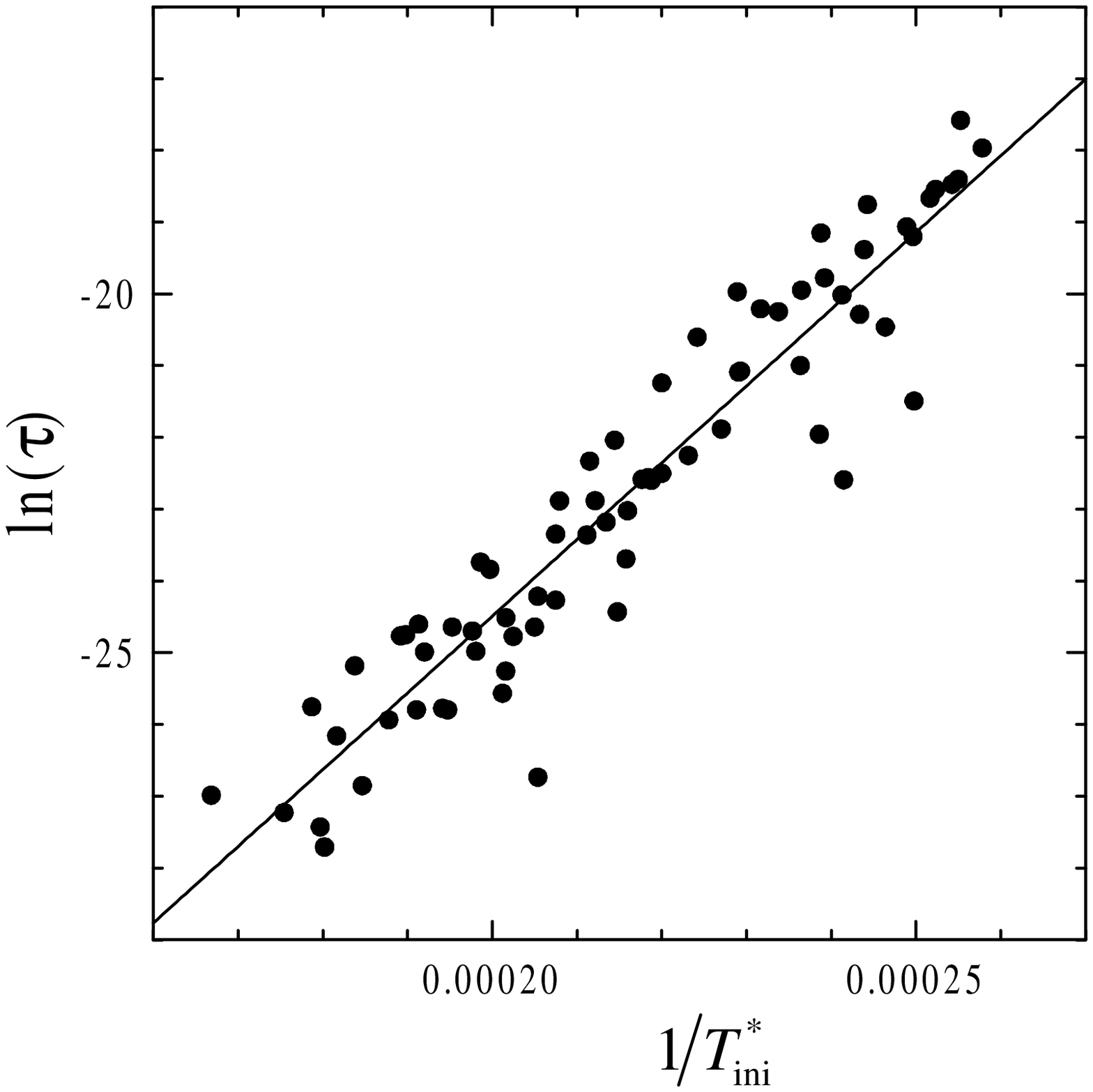}
\vskip 20mm
\centerline{Figure 3}


\newpage

\begin{thebibliography}{10}

\bibitem{1}H.W. Kroto, J.R. Heath, S.C. O'Brien et al., Nature $\mathbf{318}$, 162 (1985).
\bibitem{2}A.D. Boese, G.E. Scuseria, Chem. Phys. Lett. $\mathbf{294}$, 233 (1998).
\bibitem{3}G. S\'anchez, S. D{\fontencoding{T1}\selectfont\symbol{237}}az-Tendero, M. Alcam{\fontencoding{T1}\selectfont\symbol{237}}, F. Mart{\fontencoding{T1}\selectfont\symbol{237}}n, Chem. Phys. Lett. $\mathbf{416}$, 14 (2005).
\bibitem{4}C. Lifshitz, Int. J. Mass Spectrom. $\mathbf{198}$, 1 (2000).
\bibitem{5}J. Laskin, B. Hadas, T.D. M\"ark, C. Lifshitz, Int. J. Mass Spectrom. $\mathbf{177}$, L9 (1998).
\bibitem{6}S. Matt, O. Echt, M. Sonderegger et al., Chem. Phys. Lett. $\mathbf{303}$, 379 (1999).
\bibitem{7}S. Matt, O. Echt, P. Scheirer, T.D. M\"ark, Chem. Phys. Lett. $\mathbf{348}$, 194 (2001).
\bibitem{8}K. Hansen, O. Echt, Phys. Rev. Lett. $\mathbf{78}$, 2337 (1997).
\bibitem{9}S. Tomita, J.U. Andersen, C. Gottrup et al., Phys. Rev. Lett. $\mathbf{87}$, 073401 (2001).
\bibitem{10}S. Tomita, J.U. Andersen, K. Hansen, P. Hvelplund, Chem. Phys. Lett. $\mathbf{382}$, 120 (2003).
\bibitem{11}J.U. Andersen, E. Bonderup, K. Hansen et al., Eur. Phys. J. D $\mathbf{24}$, 191 (2003).
\bibitem{12}K. G{\fontencoding{T1}\selectfont\l}uch, S. Matt-Leubner, O. Echt et al., J. Chem. Phys. $\mathbf{121}$, 2137 (2004).
\bibitem{13}B. Concina, K. G{\fontencoding{T1}\selectfont\l}uch, S. Matt-Leubner
et al., Chem. Phys. Lett. $\mathbf{407}$, 464 (2005).
\bibitem{14}E. Kim, Y.H. Lee, Phys. Rev. B $\mathbf{48}$, 18230 (1993).
\bibitem{15}S.G. Kim, D. Tom\'anek, Phys. Rev. Lett. $\mathbf{72}$, 2418 (1994).
\bibitem{16}S. Serra, S. Sanguinetti, L. Colombo, Chem. Phys. Lett. $\mathbf{225}$, 191 (1994).
\bibitem{17}C. Xu, G.E. Scuseria, Phys. Rev. Lett. $\mathbf{72}$, 669 (1994).
\bibitem{18}E. Kim, D.-H. Oh, C.W. Oh, Y.H. Lee, Synth. Met. $\mathbf{70}$, 1495 (1995).
\bibitem{19} A. I. Podlivaev and L. A. Openov, Pis'ma Zh. Eksp. Teor. Fiz. $\mathbf{81}$, 656 (2005)
[JETP Lett. $\mathbf{81}$, 533 (2005)]; cond-mat/0506571.
\bibitem{20}C.H. Xu, C.Z. Wang, C.T. Chan, K.M. Ho, J. Phys.: Condens. Matter $\mathbf{4}$, 6047 (1992).
\bibitem{21} Yu. E. Lozovik and A. M. Popov, Usp. Fiz. Nauk $\mathbf{167}$, 751 (1997) [Phys. Usp. $\mathbf{40}$, 717 (1997)].
\bibitem{22}I. V. Davydov, A. I. Podlivaev, and L. A. Openov, Fiz. Tverd. Tela (St. Petersburg) $\mathbf{47}$, 751 (2005)
[Phys. Solid State $\mathbf{47}$, 778 (2005)]; cond-mat/0503500.
\bibitem{23}A.J. Stone, D.J. Wales, Chem. Phys. Lett. $\mathbf{128}$, 501 (1986).
\bibitem{24}P.A. Marcos, J.A. Alonso, M.J. L\'opez, J. Chem. Phys. $\mathbf{123}$, 204323 (2005).
\bibitem{25}K.R. Lykke, Phys. Rev. A $\mathbf{52}$, 1354 (1995).
\bibitem{26}C.E. Klots, Z. Phys. D $\mathbf{20}$, 105 (1991).
\bibitem{27}J.V. Andersen, E. Bonderup, K. Hansen, J. Chem. Phys. $\mathbf{114}$, 6518 (2001).
\bibitem{28}P.A. Marcos, J.A. Alonso, A. Rubio, M.J. L\'opez, Eur. Phys. J. D $\mathbf{6}$, 221 (1999).
\bibitem{29}K. Hansen, E. E. B. Campbell, and O. Echt, Int. J. Mass Spectrom. $\mathbf{252}$, 72 (2006).

\end{thebibliography}
\end{document}